\documentclass[12pt,pre,aps,showpacs,preprint]{revtex4}

\usepackage{graphics}
\usepackage{graphicx}
\usepackage{amsfonts}
\usepackage{textcomp}
\usepackage{amssymb}
\usepackage{amsmath}

\begin{document}

\title{Novel Features Arising in Maximally Random Jammed Packings of Superballs}

\author{Y. Jiao$^1$, F. H. Stillinger$^2$ and S. Torquato$^{2,3,4,5}$}


\affiliation{$^1$Department of Mechanical and Aerospace
Engineering, Princeton University, Princeton New Jersey 08544,
USA}



\affiliation{$^2$Department of Chemistry, Princeton University,
Princeton New Jersey 08544, USA}





\affiliation{$^3$Program in Applied and Computational Mathematics,
Princeton University, Princeton New Jersey 08544, USA}

\affiliation{$^5$Princeton Center for Theoretical Physics,
Princeton University, Princeton New Jersey 08544, USA}

\affiliation{$^4$School of Natural Sciences, Institute for
Advanced Study, Princeton NJ 08540}

\begin{abstract}

Dense random packings of hard particles are useful models of granular 
media and are closely related to the structure of nonequilibrium low-temperature 
amorphous phases of matter. Most work has been done for random jammed packings of spheres, 
and it is only recently that corresponding packings of nonspherical particles (e.g., ellipsoids) 
have received attention. Here we report a study of the maximally random jammed (MRJ) packings 
of binary superdisks and monodispersed superballs whose shapes are defined by 
$|x_1|^{2p}+\cdots+|x_d|^{2p}\le1$ with $d=2$ and $3$, respectively, where 
$p$ is the deformation parameter with values in the interval $(0, \infty)$. As $p$ increases 
from zero, one can get a family of both concave ($0<p<0.5$) and convex ($p\ge 0.5$) particles with 
square symmetry ($d=2$), or octahedral and cubic symmetry ($d=3$). In particular, 
for $p=1$ the particle is a perfect sphere (circular disk) and for $p\rightarrow \infty$ 
the particle is a perfect cube (square).   
We find that the MRJ densities of such packings
increase dramatically and nonanalytically 
as one moves away from the circular-disk and sphere point ($p=1$). Moreover, 
the disordered packings are hypostatic, i.e., the average number of contacting 
neighbors is less than twice the total number of degrees of freedom per particle,
and the packings are mechanically stable. 
As a result, the local arrangements of particles are necessarily nontrivially 
correlated to achieve jamming. We term such correlated structures ``nongeneric''. 
The degree of ``nongenericity'' of the packings 
is quantitatively characterized by determining the fraction of local coordination structures 
in which the central particles have fewer contacting neighbors than average. 
We also show that such seemingly ``special'' packing configurations are 
counterintuitively not rare. As the anisotropy of the particles increases,
the fraction of rattlers decreases while the minimal orientational order as measured 
by the cubatic order metric increases. 
These novel characteristics result from the unique rotational symmetry breaking manner of the particles, 
which also makes the superdisk and superball packings 
distinctly different from other known nonspherical hard-particle packings.

\end{abstract}

\pacs{61.50.Ah, 05.20.Jj}

\maketitle

\section{Introduction}

Particle packing problems, such as how to fill a volume with given solid objects 
as densely as possible, are among the most ancient and persistent problems 
in science and mathematics. A \textit{packing} is a large collection of 
non-overlapping solid objects (particles) in $d$-dimensional Euclidean space 
$\mathbb{R}^d$. The packing density $\phi$ is defined as the fraction of space 
$\mathbb{R}^d$ covered by the particles. Dense ordered and random packings of 
nonoverlapping (hard) particles have been employed to understand the 
equilibrium and non-equilibrium structure of a variety many-particle systems, 
including crystals, glasses, heterogeneous materials and granular media 
\cite{Zallen,Edwards,Chaikin,SalBook}. 
Packing problems in dimensions higher than three attract current interest 
for retrieving stored data transmitted through a noisy channel \cite{Shannon,Conway,Sc08,Co09}. 

The packings of congruent hard spheres in $\mathbb{R}^3$ have been intensively studied 
since despite the simplicity they exhibit rich packing characteristics.
It is only recently that the densest packings with $\phi_{max} = \pi/\sqrt{18} \approx 0.74$, 
realized by the face-centered cubic lattice and its stacking variants, have been proved \cite{Hales}.  
In addition, three-dimensional random packings can be prepared both experimentally and numerically
with a relatively robust density $\phi \approx 0.64$ \cite{Scott, SalMRJ}.
The term \textit{random close packing} (RCP) \cite{nature_notes}, widely used to designate 
 the ``random" packing with the highest achievable density,
is ill-defined since random packings can be obtained as the system becomes more ordered and 
a definition of randomness has been lacking. A more recent concept that has been suggested to replace RCP is 
that of the maximally random jammed (MRJ) state \cite{SalMRJ}, corresponding to 
the most disordered among all jammed (mechanically stable) packings. 
A jammed packing is one in which the particle positions and orientations are fixed by 
the impenetrability constraints and boundary conditions  \cite{SalJam}.
It has been established that the MRJ state for spheres in $\mathbb{R}^3$ has a density of $\phi \approx 0.637$,
as obtained by a variety of different order metrics \cite{SalJam,SalMRJ2}. This 
density value is consistent with what has traditionally asscoiated with RCP in three dimensions.

It has been argued in the granular materials literature that large 
disordered jammed (MRJ) packings of hard frictionless spheres are \textit{isostatic} \cite{Al98, Ed99}, 
meaning that the total number of interparticle contacts (constraints) 
equals the total number of degrees of freedom of system and that 
all of the constraints are (linearly) independent. This implies that the 
average number of contacts per particle $Z$ is equal to twice the number of degrees 
of freedom per particle $f$ (i.e., $Z=2f$), in the limit as the number of 
particles gets large. This prediction has been verified computationally 
with very high accuracy \cite{Aleks_g2, Ohern}. On the other hand, 
a packing is \textit{hypostatic} if it is mechanically stable (i.e., jammed) while 
the number of constraints is smaller than the number of degrees of freedom. For 
large packings, this is equivalent to the inequality $Z<2f$. It has been shown that a jammed sphere 
packing can be not hypostatic \cite{Aleks_g2}.


It is also of great practical and fundamental interest to understand 
the organizing principles of dense packings of nonspherical 
particles \cite{SpheroCylinI, SpheroCylinII,  AlexPRLDense, Frenkel, Burton, Ken,Jiao08PRL, Jiao09PRE, Sal09Nat, Sal09PRE}. 
The effect of asphericity is an important feature to include 
on the way to characterizing more completely real dense granular 
media as well as low-temperature states of matter. Another important application
relates to supramolecular chemistry \cite{supra} of organic compounds
whose molecular constituents can possess many different types
of group symmetries \cite{Ki73}. Such systems can be approximated 
by nonspherical hard particles with the same group symmetries.  

Recently, MRJ packings of three-dimensional ellipsoids \cite{AlexSci, AlexPREJam} 
have been studied. In particular, it was found that the density $\phi$ and the average 
coordination number $Z$ (the average number of touching neighbors per particle) increase rapidly, in a 
cusp-like manner, as asphericity is introduced from the sphere point. 
The density $\phi$ reaches a maximum at a critical 
aspect ratio $\alpha^*$ \cite{aspect} and then begins to decrease; while $Z$ is increasing monotonically 
until it attains the plateau value for all $\alpha$ beyond $\alpha^*$. In addition, $Z$ is 
always smaller than twice the number of degrees of freedom per particle $f$ 
with its plateau value slightly below $2f$ (for an ellipse $f=3$ and for an ellipsoid $f=6$). 
In other words, the packings are \textit{hypostatic} \cite{AlexPREJam}. 
The characteristics of MRJ ellipsoid packings are 
distinctly different from their densest crystalline counterpart \cite{AlexPRLDense}, 
in which $\phi$ increases smoothly as one moves away from the sphere point, and 
reaches a plateau value of $0.7707...$ for $\alpha > \sqrt{3}$ (oblate spheroids) and 
$\alpha < 1/\sqrt{3}$ (prolate spheroids).

\begin{figure}
\begin{center}
$\begin{array}{c}
\includegraphics[height=3.0cm, keepaspectratio]{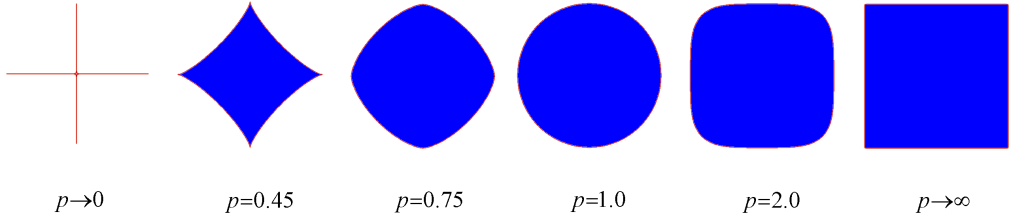} \\
\end{array}$
\end{center}
\caption{(color online). Superdisks with different values of the deformation parameter $p$.}
\label{fig1}
\end{figure}
\begin{figure}
\begin{center}
$\begin{array}{c}
\includegraphics[height=3.5cm, keepaspectratio]{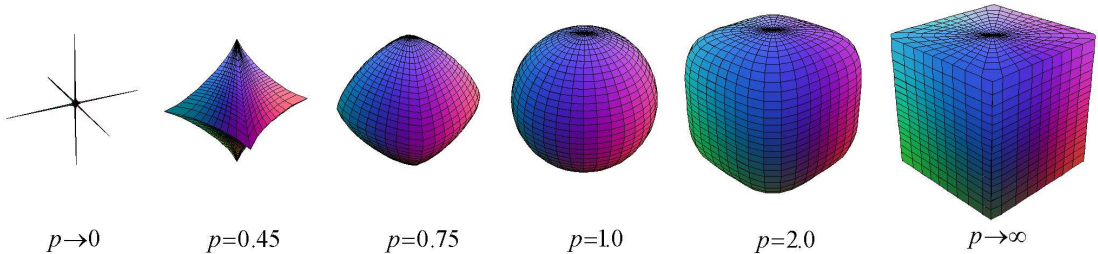} \\
\end{array}$
\end{center}
\caption{(color online). Superballs with different values of the deformation
parameter $p$.} \label{fig2}
\end{figure}

In Refs.~\cite{Jiao08PRL} and \cite{Jiao09PRE}, we studied dense and maximally 
dense packings of \textit{superballs}, a family of nonspherical particles with versatile shapes.
In particular, a $d$-dimensional superball is a centrally symmetric body in 
$\mathbb{R}^d$ occupying the region

\begin{equation}
|x_1|^{2p}+|x_2|^{2p}+\cdots+|x_d|^{2p} \le 1,
\end{equation}

\noindent where $x_i$ $(i=1,\ldots,d)$ are Cartesian coordinates
and $p \ge 0$ is the \textit{deformation parameter}, which
indicates to what extent the particle shape has deformed from that
of a $d$-dimensional sphere ($p=1$).
Henceforth, the terms \textit{superdisk} and \textit{superball} 
will be our designations
for the two-dimensional ($d=2$) and three-dimensional ($d=3$)
cases, respectively. A superdisk possesses square 
symmetry, as $p$ moves away from unity,
two families of superdisks can be obtained,
with the symmetry axes rotated 45 degrees with respect to each
other; when $p<0.5$, the superdisk is concave (see Fig.~\ref{fig1}).
A superball can possess two types of 
shape anisotropy: cube-like shapes (for $p>1$) 
and octahedron-like shapes (for $0<p<1$) with a shape change 
from convexity to concavity as $p$ passes downward through $0.5$ (see Fig.~\ref{fig2}).

Optimal packings of congruent superdisks and superballs apparently are realized by 
certain Bravais lattices possessing symmetries consistent with 
those of the particles  \cite{Jiao08PRL, Jiao09PRE, Sal09PRE}. Even these crystalline packings exhibit 
rich characteristics that are distinctly different from other known 
packings of nonspherical particles. 
For example, we found that the maximal density $\phi_{max}$ as a function of $p$ 
at $p=1$ (the sphere or circular-disk point) is nonanalytic 
and increases dramatically as $p$ moves away from unity. In addition, 
we have discovered two-fold degenerate maximal 
density states for square-like superdisks, and both cube-like and 
octahedron-like superballs.


In this paper, we generate both packings of binary superdisks in $\mathbb{R}^2$ 
and monodisperse superballs in $\mathbb{R}^3$ 
that represent the maximally random jammed (MRJ) state of these particles, 
using a novel event-driven molecular dynamics algorithm \cite{Do05a, Do05b} 
and investigate their characteristics. 
For both superdisks and superballs, we find that 
the corresponding density $\phi$ and the average contact number $Z$ 
increase rapidly, in a cusp-like manner, as the particles 
deviate from perfect circular disks and spheres, respectively. 
In particular, we find that the MRJ packing density 
$\phi$ increases monotonically as $p$ moves away from unity, 
and shows no signs of a plateau even for large $p$ values.
This is to be contrasted with the case of ellipsoids for
which the packing density reaches a maximum as the aspect ratio ratio increases from
its sphere-point value and then begins to decrease 
as the aspect ratio grows beyond that associated with the density-maximum value. 

Moreover, we find that $Z$ for superdisk and  superball packings reach its associated
plateau value at relatively small asphericity deviations (i.e., $|p-1|$) 
and the packings remain \textit{hypostatic} for all 
values of $p$ examined. By ``hypostatic'', we mean that the $Z$ is smaller than 
twice the number of degrees of freedom per particle, compared 
to random ellipsoid packings where the plateau value 
of $Z$ is only slightly below $2f$. Therefore, to achieve jamming, 
the local particle arrangements are necessarily correlated in 
a nontrivial way. We call such correlated structures ``nongeneric'' \cite{ftn_degenerate}.
We quantify the degree of ``nongenericity'' of the packings 
by determining the fraction of local coordination configurations in which 
the central particles have fewer contacting neighbors than average $Z$. 
We also show that such ``nongeneric'' configurations 
are not rare, which is a rather counterintuitive conclusion.
In addition, we find that although the rapid increase of density 
is unrelated to any observable translational order, 
the orientational order (e.g., the cubatic order 
parameter \cite{Cubatic}) increases as $p$ 
moves away from unity.  
These packing characteristics, which are distinctly different 
from that of the MRJ packings of ellipsoids, 
are due to  the unique way in which rotational symmetry is broken in 
superdisk and superball packings. 

The rest of the paper is organized as follows: In Sec. II, 
we briefly describe the simulation techniques and the obtained 
packings. In Sec. III, we provide a detailed analysis of 
the novel packing characteristics. In Sec. IV, we make 
concluding remarks.


\section{Maximally Random Jammed Packings via Computer Simulation}

We use an event-driven molecular dynamics packing algorithm recently 
developed by Donev, Torquato and Stillinger \cite{Do05a, Do05b} 
(henceforth, referred to as the DTS algorithm) to 
generate MRJ packings of convex superdisks in two dimensions and superballs in three dimensions.
The DTS algorithm generalizes the Lubachevsky-Stillinger (LS)
sphere-packing algorithm \cite{LSpacking} to the case of other
centrally symmetric convex bodies (e.g., ellipsoids and
superballs). Initially, small particles 
are randomly distributed and randomly
oriented in the simulation box (fundamental cell) with periodic
boundary conditions and without any overlap. The particles are
then given translational and rotational velocities randomly and
their motion followed as they collide elastically and also expand
uniformly with an expansion rate $\gamma$, while the fundamental cell deforms to better
accommodate the configuration. After some time, a jammed state
with a diverging collision rate is reached and the
density reaches a local maximum value. To generate random 
jammed packings, initially large $\gamma$ are employed to prevent 
the system following the equilibrium branch of the phase diagram 
that leads to crystallization. Near the jamming point, 
sufficiently small expansion rate is necessary for the particles 
to establish contacting neighbor networks and to form a truly 
jammed packing. 
 On the basis of our experience with spheres \cite{Aleks_g2} and ellipsoids \cite{AlexSci},
 we believe that our algorithm with rapid particle expansion produces final 
states that represent the MRJ state well.
Here we use the largest possible initial $\gamma \in (0.1 - 0.5)$ that is numerically feasible to 
ensure the generated superdisk and superball packings are maximally random jammed. 
We mainly focus on superdisks and superballs with deformation 
parameters $p$ within the range $0.85 - 3.0$, since extreme values
of $p$ associated with polyhedron-like shapes present
numerical difficulties.

All of the generated packings used in the subsequent
analyses are verified to be at least collectively jammed using
an ``infinitesimal shrinkage'' method \cite{SalJam}, i.e., the particles in
the packing are shrunk by a very small amount   \cite{shrinkage} and given random velocities. If
no significant structural changes occur after the system
``relaxes'' after a sufficiently long enough time, the packing is considered
to be collectively jammed. It is well established that
the ``infinitesimal shrinkage'' method is
robust, i.e., it always gives the same results for sphere
packings as those obtained from a rigorous linear programming
jamming test algorithm provided the amount of shrinkage from the jammed
state is sufficiently small for a given number of particles
within the periodic cell \cite{Do04}.

\subsection{Binary Mixtures of MRJ Superdisks}


\begin{figure}
\begin{center}
$\begin{array}{c@{\hspace{1.75cm}}c}
\includegraphics[height=6.0cm, keepaspectratio]{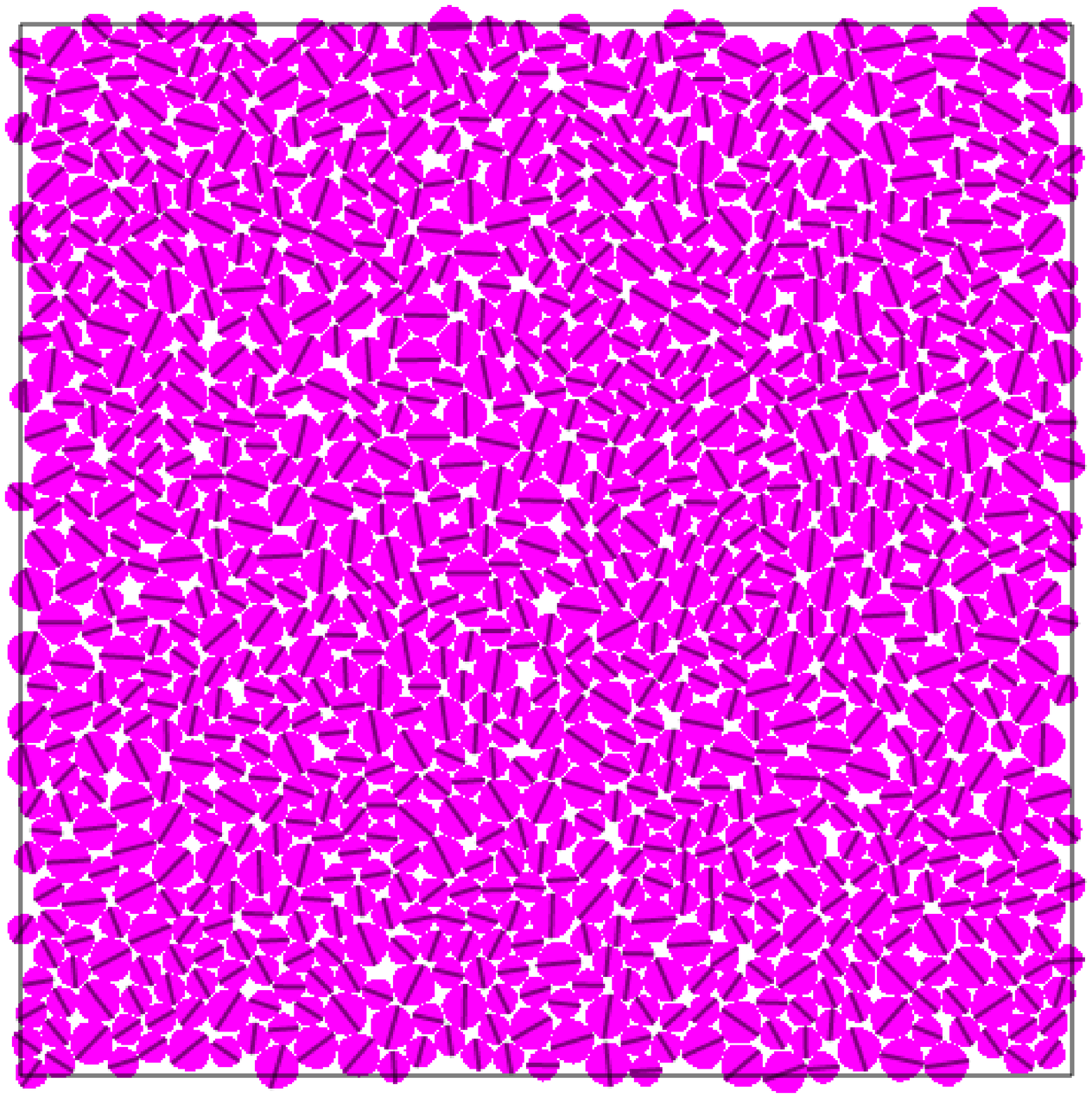} &
\includegraphics[height=6.0cm, keepaspectratio]{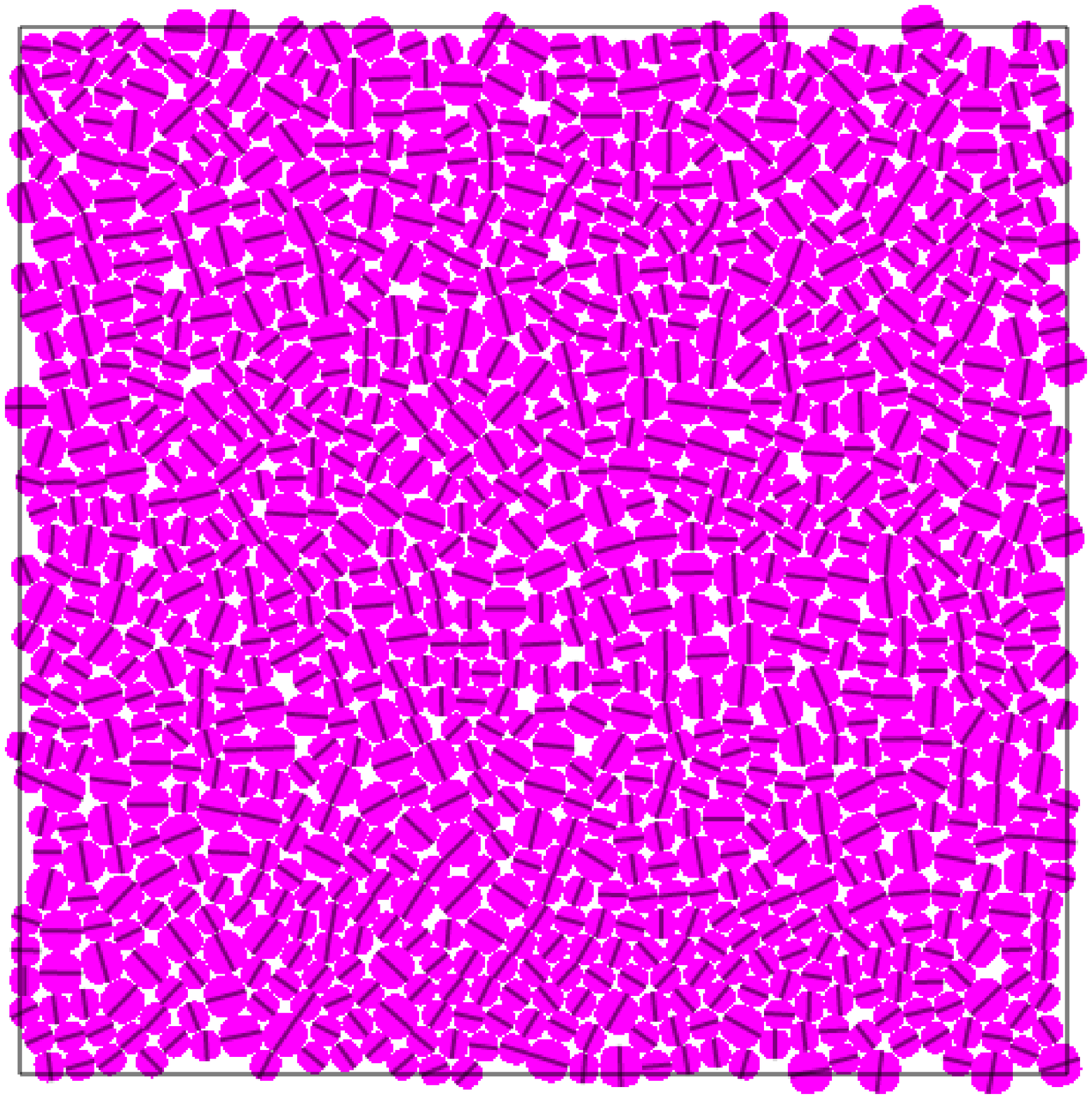} \\
\mbox{(a) $p=0.85$} & \mbox{(b) $p=1.5$}
\end{array}$
\end{center}
\caption{(color online). Typical configurations of MRJ packings of binary superdisks 
with different values of the deformation parameter $p$. The chords 
show one of the symmetry axes of the superdisks.}
\label{Superdisks}
\end{figure}

\begin{figure}
\begin{center}
$\begin{array}{c}
\includegraphics[height=7.0cm, keepaspectratio]{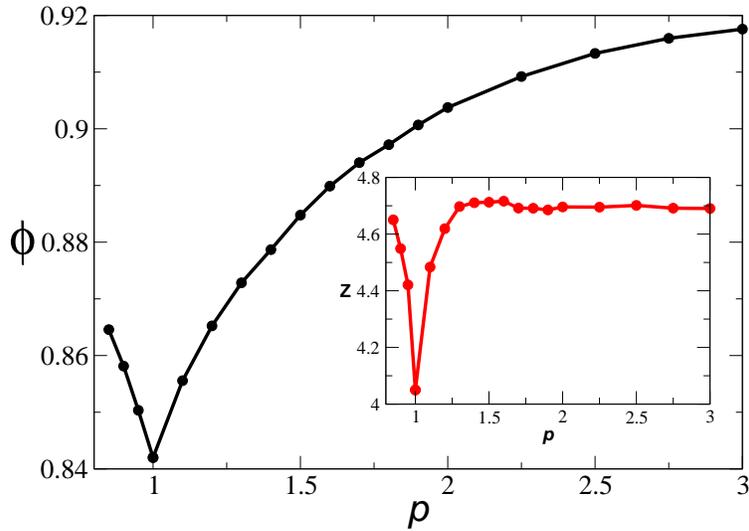}
\end{array}$
\end{center}
\caption{The density of MRJ packings of binary superdisks as a function 
of $p$. Insert: the average contact number $Z$ as a function of $p$.}
\label{2D_rho_Z}
\end{figure}

In two dimensions, we study MRJ packings of 
a specific family of binary superdisk mixtures in which 
the size ratio is $\kappa = 1.4$ and the molar 
ratio is $\beta = 1/3$. The size ratio $\kappa$ is defined as the ratio of the 
diameter of large superdisks over that of the small superdisks; and
the molar ratio $\beta$ is defined as the number large superdisks over the number of small superdisks.
We do not use monodispersed 
superdisk systems here because they are easily crystallized into ordered 
packings \cite{Jiao08PRL}. For $p=1$, one obtains the binary 
circular-disk system which has been intensively studied as a 
prototypical glass former \cite{binarydisk}. Typical jammed packing 
configurations are shown in Fig.~\ref{Superdisks}. The density $\phi$ and 
the average contact number per particle $Z$ as a function of $p$ 
are shown in Fig.~\ref{2D_rho_Z}, which reveals that the 
initial rapid increases of $\phi$ and $Z$ are linear in $|p-1|$ \cite{slope2d}. 
The density $\phi$ increases monotonically as $p$ moves away from unity 
and shows no signs of a plateau, even for relatively large $p$. 
In addition, $\phi$ quickly surpasses the density of 
the optimal binary circular-disk packing associated with the 
size and molar ratios employed here, which contains phase-separate 
regions of triangular lattice packings of different sized circular disks \cite{binarydisk}.
The quantity $Z$ quickly reaches its plateau value 
$Z^* \approx 4.7$ at $p\approx 1.3$, which is smaller 
than $2f= 6$, indicating the packings are hypostatic. 
The cubatic order parameter $P_4$ is defined as 
$P_4 = \langle(35\cos^4\theta - 30\cos^2\theta+3)/8\rangle$, where $\theta$ is 
the angle between the particle axis and the director, along which the 
principle axes of the particles have maximum mutual alignment \cite{Cubatic}.
The measured cubatic order parameter is $P_4 \approx 0.06$ to $0.32$ 
with the tendency to increase as $|p-1|$ grows. 

\subsection{MRJ Packings of Monodisperse Superballs}


\begin{figure}
\begin{center}
$\begin{array}{c@{\hspace{1.75cm}}c}
\includegraphics[height=6.5cm, keepaspectratio]{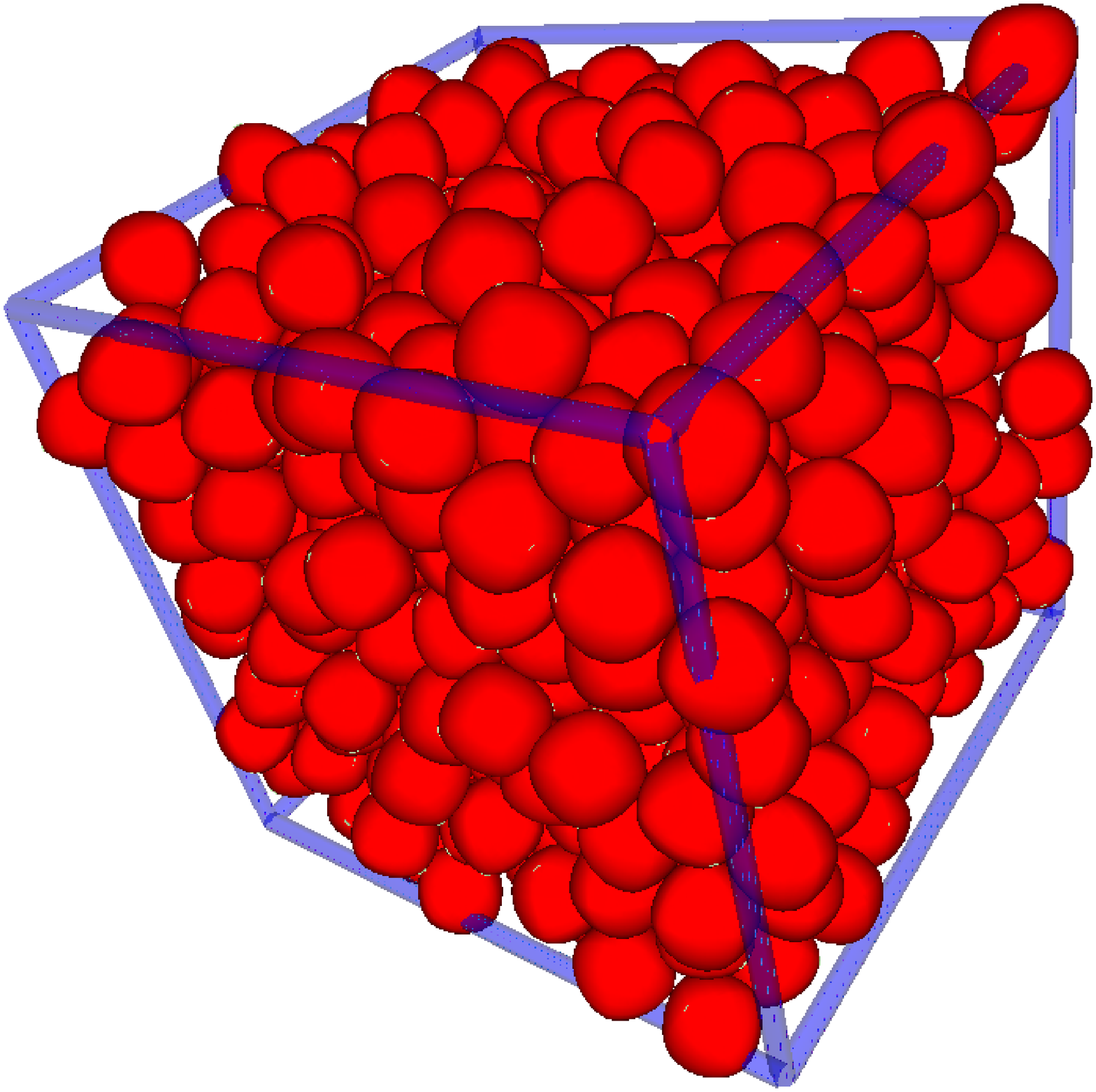} &
\includegraphics[height=6.5cm, keepaspectratio]{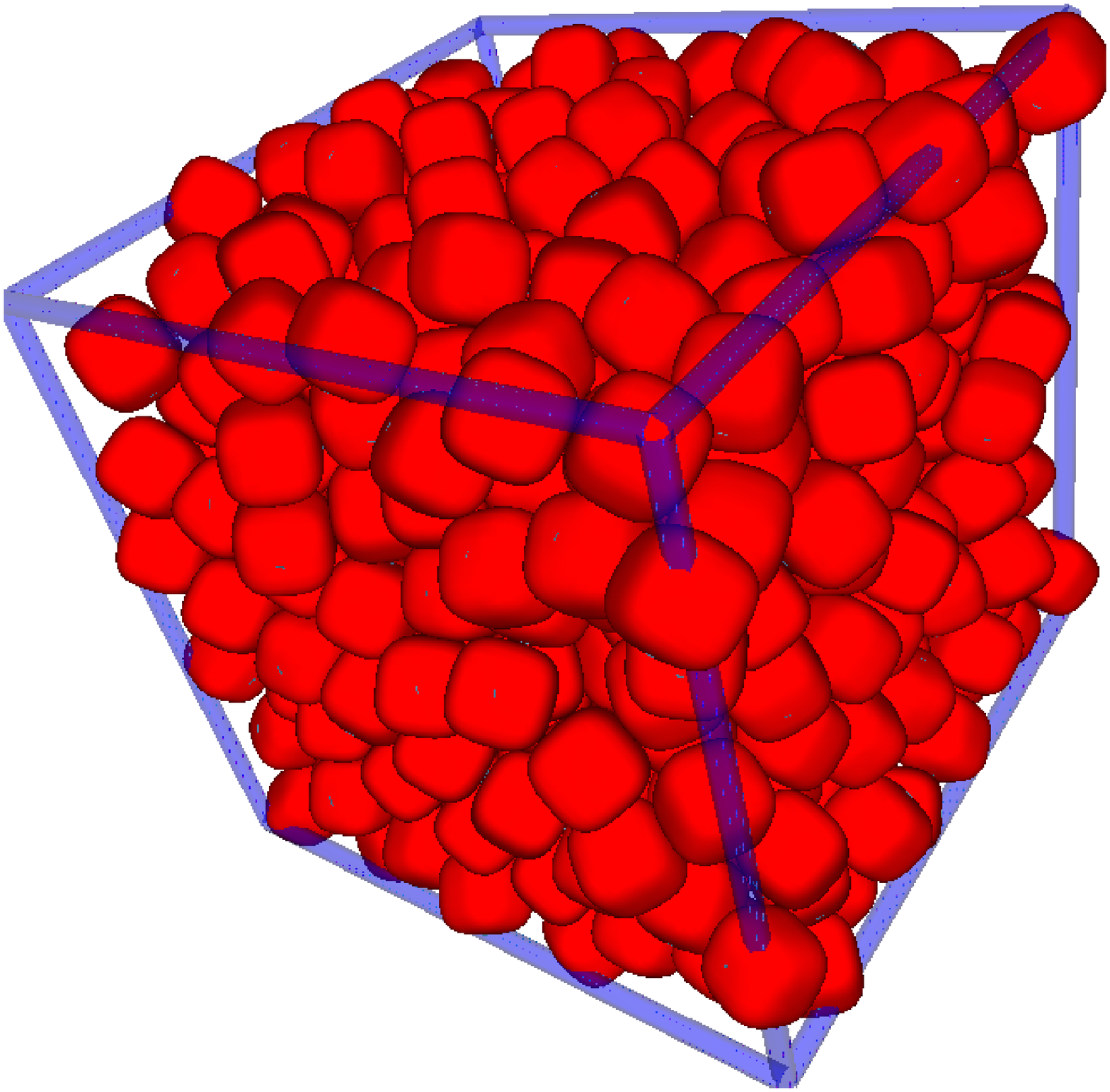} \\
\mbox{(a) $p=0.85$} & \mbox{(b) $p=1.5$}
\end{array}$
\end{center}
\caption{(color online). Typical configurations of MRJ packings of superballs 
with different values of the deformation parameter $p$.}
\label{Superballs}
\end{figure}

\begin{figure}
\begin{center}
$\begin{array}{c}
\includegraphics[height=7.0cm, keepaspectratio]{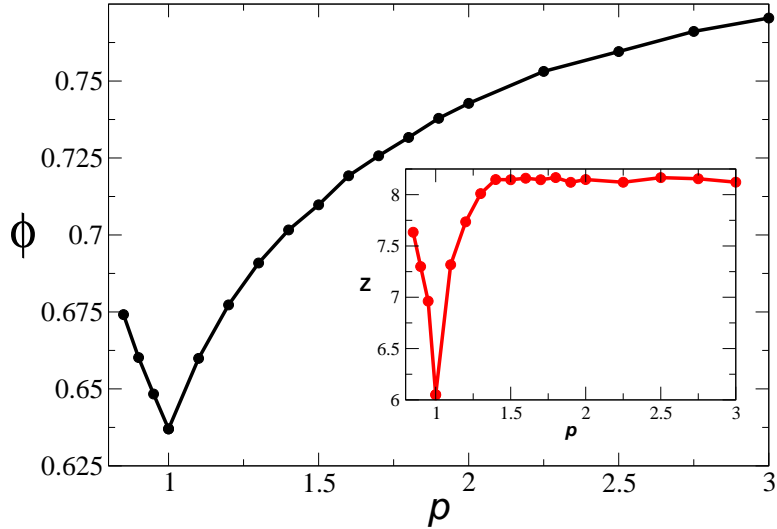}
\end{array}$
\end{center}
\caption{The density of MRJ packings of superballs as a function 
of $p$. Insert: the average contact number $Z$ as a function of $p$.}
\label{3D_rho_Z}
\end{figure}

In three dimensions, monodispersed superballs can be easily 
compressed into a jammed random packing due to geometrical 
frustration (i.e., the densest local particle arrangement 
cannot tile space). Typical jammed packing 
configurations are shown in Fig.~\ref{Superballs}. The quantities $\phi$ and $Z$ as a function of $p$ 
are shown in Fig.~\ref{3D_rho_Z}, respectively. As in two dimensions, 
$\phi$ and $Z$ increase rapidly, in a cusp-like manner \cite{slope3d}, as the particles 
deviate from perfect sphere. $\phi$ increases monotonically as $p$ moves 
away from unity, quickly goes beyond the optimal sphere packing density 
and shows no signs of plateau, even for relatively large $p$ values.
The contact number per particle $Z$ reaches its plateau value $Z^* \approx 8.15$ at $p \approx 1.4$, 
which is significantly smaller than $2f= 12$, 
indicating that the packings are hypostatic. 
The measured cubatic order parameter is $P_4 \approx 0.03$ to $0.21$, 
which increases with $|p-1|$. 

\section{Packing Characteristics}

\subsection{Rattlers}

MRJ packings generated in both two and three dimensions 
contain a small fraction of rattlers, i.e., particles that 
can wander freely within cages formed by their jammed non-rattling neighbors. 
When $p$ is close to unity, the fraction of rattlers is approximately 
$2.6\%$ and $1.2\%$ for two and three dimensions, respectively. 
As $p$ moves away from unity, the fraction of rattlers decreases 
quickly and practically vanishes for large $p$ (e.g., $p>2.75$). 
This behavior results from the increasing protuberance of the particle shape, 
which makes it more difficult to form isotropic cages and also 
requires more average contacts per particle to achieve jamming. 
Note that rattlers are excluded when reporting average 
contact numbers in the following discussion.

\subsection{Packing Density}

The rapid increase of the density is mainly due to 
the broken rotational symmetry of the particles. In particular, 
the cubic-like (square-like) and octahedral-like particles are 
more efficient to cover the space than spheres (circular disks), 
i.e., near the jamming point the particles 
can rotate to accommodate the neighbors by orienting the 
``far corners'' to fill the available gaps and thus cover
more space. For small values of $p$, the increase 
in $\phi$ is also attributed to the expected increase in the 
number of contacting neighbors per particle, which means 
locally more particles can be packed in a given volume. 
The manner in which rotational symmetry is broken in superball packings is distinctly 
different from that in ellipsoid packings. For example, the asphericity $\gamma$ \cite{Sal09Nat, Sal09PRE}, 
defined as the ratio of the radii of circumsphere and insphere 
of a nonspherical particle, is always bounded and close to 
unity for all values of $p$ for superballs, while it can 
increase without limit as the largest aspect ratio $\alpha$ grows for 
ellipsoids. For very elongated or flake-like ellipsoids with large 
aspect ratios, the effect of 
a very anisotropic exclusion volume becomes dominant and causes 
the density of random ellipsoid packings to decrease. 
By contrast, the shape of superballs becomes more 
efficient in filling space as the deformation parameter deviates 
more from unity and thus results in a monotonically increasing 
density. The nonanalyticity of $\phi$ at $p=1$ is also associated 
with the broken symmetry of superdisks and superballs. 
This nonanalytical behavior has also been observed in the 
optimal packings of these particles realized by various 
Bravais lattices \cite{Jiao08PRL, Jiao09PRE}. 
This stands in contrast to the densest known ellipsoid packings,
which are periodic packings with a two-particle basis possessing 
a smooth initial increase of $\phi_{max}$ as 
the aspect ratio moves away from unity \cite{AlexPRLDense}.

\subsection{Hypostaticity and Nongeneric Local Structures}

There have been conjectures \cite{Al98, Ed99} that frictionless random packings 
have just enough constraints to completely statically define the 
system (i.e., it is \textit{isostatic}), i.e., for large packings, one has $Z = 2f$.
It has been shown both experimentally and computationally that 
although the isostatic conjecture \cite{Al98} holds for large sphere packings \cite{Be90, Do04},
it is generally not applicable to nonspherical 
particles, such as ellipsoids \cite{AlexSci, AlexPREJam}.
It was found that even for ellipsoids with large aspect ratios, 
$Z$ is still slightly below $2f$ \cite{AlexSci}.

\begin{figure}
\begin{center}
$\begin{array}{c@{\hspace{1.0cm}}c}
\includegraphics[height=7.25cm, keepaspectratio]{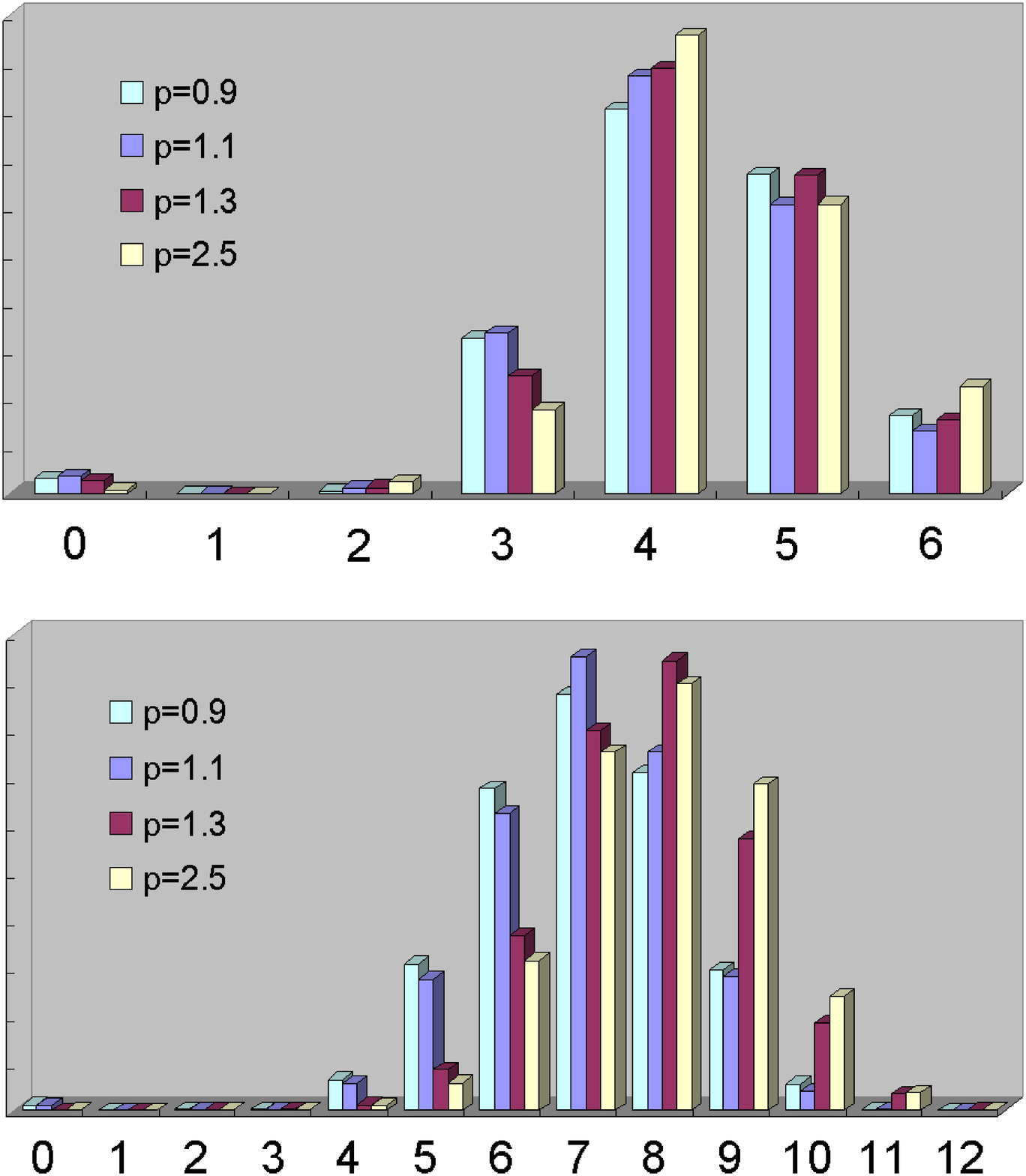} &
\includegraphics[height=7.25cm, keepaspectratio]{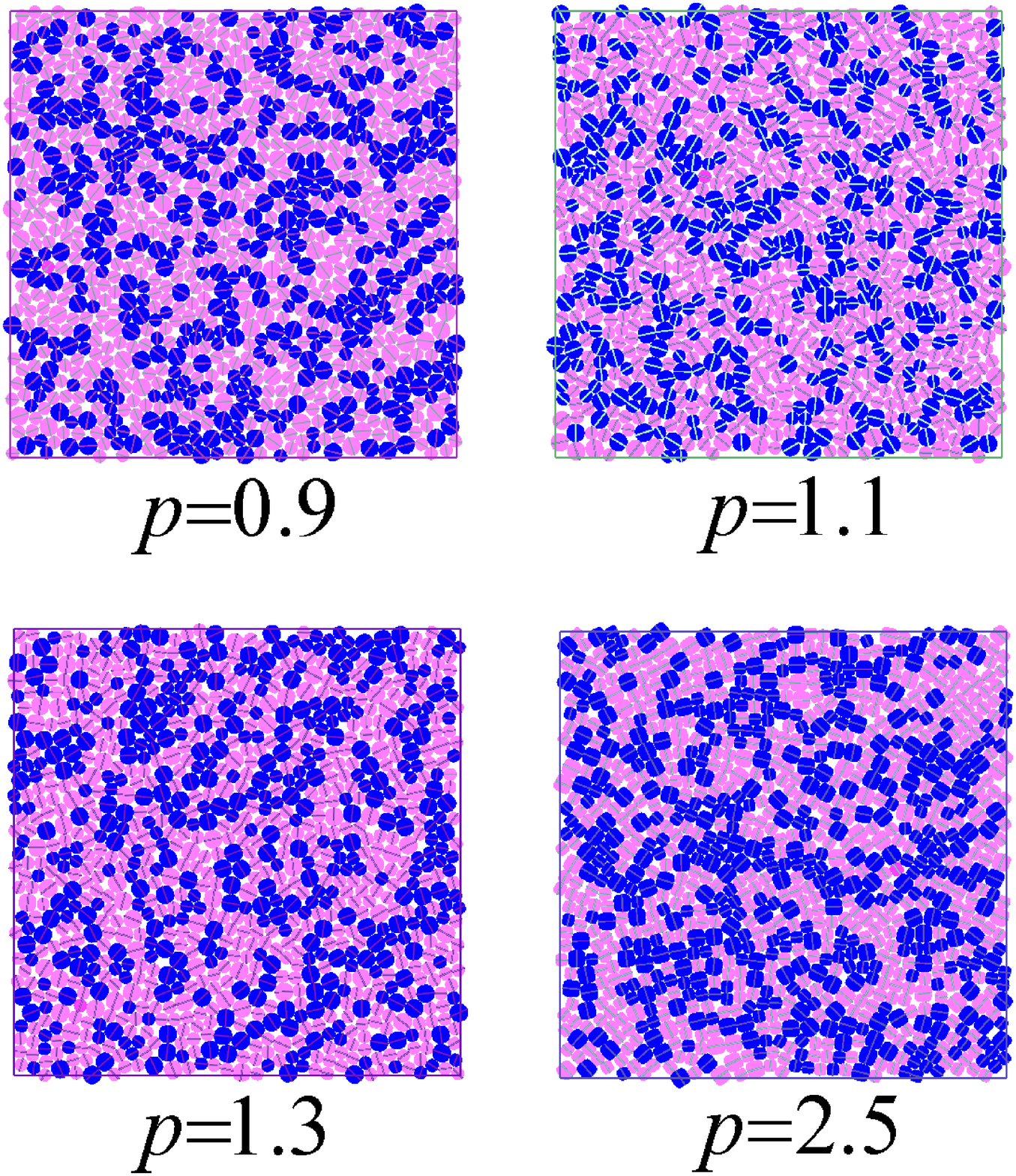} \\
\mbox{(a)} & \mbox{(b)}
\end{array}$
\end{center}
\caption{(color online). (a) Distribution of contact numbers for different 
$p$ values for MRJ packings of superdisks (upper panel) and superballs (lower panel). 
(b) Local packing structures with more contacts than average (shown in blue) 
and those with less contacts than average (shown in pink) 
in two-dimensional superdisk packings for different $p$ values.}
\label{Generic}
\end{figure}

Here we observe that in MRJ  packings of superdisks and 
superballs $Z$ is significantly smaller than $2f$ for 
all values of $p$ examined, i.e., the packings are significantly \textit{hypostatic}. 
The hypostatic packings result from the competition 
between $f_T$ translational and $f_R$ rotational degrees of freedom of the particles 
($f=f_R+f_T$) in developing the contacting networks close to the jamming point. 
In particular, although it is true that to constrain the 
translational degrees of freedom each particle needs at least $2f_T$ 
contacts, rotational degrees of freedom can be blocked with less than $2f_R$ 
additional contacts per particle if the curvatures at the 
contacting points are sufficiently small \cite{AlexPREJam}. 
In addition, due to the relatively small asphericity $\gamma$ of 
superdisks and superballs, there is little reason to expect 
the rotational motions of these particles (especially those 
with $p$ close to unity) would be frozen 
even when they are translational trapped and may only 
rattle inside small ``cages'' formed by their neighbors. 
Near the jamming point, it is expected that the particles 
can rotate significantly \cite{rotation} until the actual jamming point 
is reached, at which rotational jamming will also 
come into play, and rotational degrees of freedom are frozen with 
the number of additional contacts much less than $2f_R$. 
This is in contrast to hypostatic MRJ packings of ellipsoids 
with large aspect ratios, for which the translational and 
rotational degrees of freedom are on the same footing and, thus, the average 
contact number per particle is only slightly below twice the number of 
total degrees of freedom. 

Furthermore, the local geometry of the MRJ packings is necessarily 
nontrivially correlated (nongeneric), i.e., all the normal vectors at the points of 
contact for a particle should intersect at a common point to 
achieve torque balance and block rotations.   
In light of the isostatic conjecture, the local packing structures are less nongeneric when 
they possess larger contact numbers so that the constraining neighbors are less 
correlated. The truly generic local packing structures 
should have $Z = 2f$ per particle, for which the constraining neighbors could be 
completely uncorrelated. 
To characterize the ``nongenericity'' of the packings, 
we compute $G_{ng}$, the fraction of local structures composed 
of particles with less contacts $Z_{local}$ than average $Z_{average}$, i.e.,

\begin{equation}
G_{ng} = \frac{N(Z_{local}\le Z_{average})}{N_{total}}.
\end{equation}
 
A larger $G_{ng}$ indicates a larger degree of nongenericity.
We find $G_{ng}$ is approximately 0.65 in two dimensions and 0.78 
in three dimensions when $p$ is close to unity, which quickly 
decreases and plateaus at 0.6 and 0.68, respectively as $p$ 
increases. Figure~\ref{Generic} shows the distribution of contact numbers 
for different $p$ values and the topology of the local 
structures contributing to $G_{ng}$. It can be seen that as $p$ moves away from 
unity, the distributions become more skewed as the means shift 
to larger $Z$. Moreover, the subset of particles associated with the 
nongeneric structures do not percolate.
We do not observe any tendency of increasing $Z$ even 
for the largest $p$ values that are computationally feasible 
and we expect that MRJ packings in the cubic limit 
are also hypostatic. 
It is noteworthy that isostatic random packings of 
superdisks and superballs are difficult to construct, 
since achieving isostaticity requires $Z = 2f = 12$ 
which is necessarily associated with translational 
crystallization \cite{ellipse}.

\begin{figure}
\begin{center}
$\begin{array}{c@{\hspace{1.0cm}}c@{\hspace{1.0cm}}c}
\includegraphics[height=4.0cm, keepaspectratio]{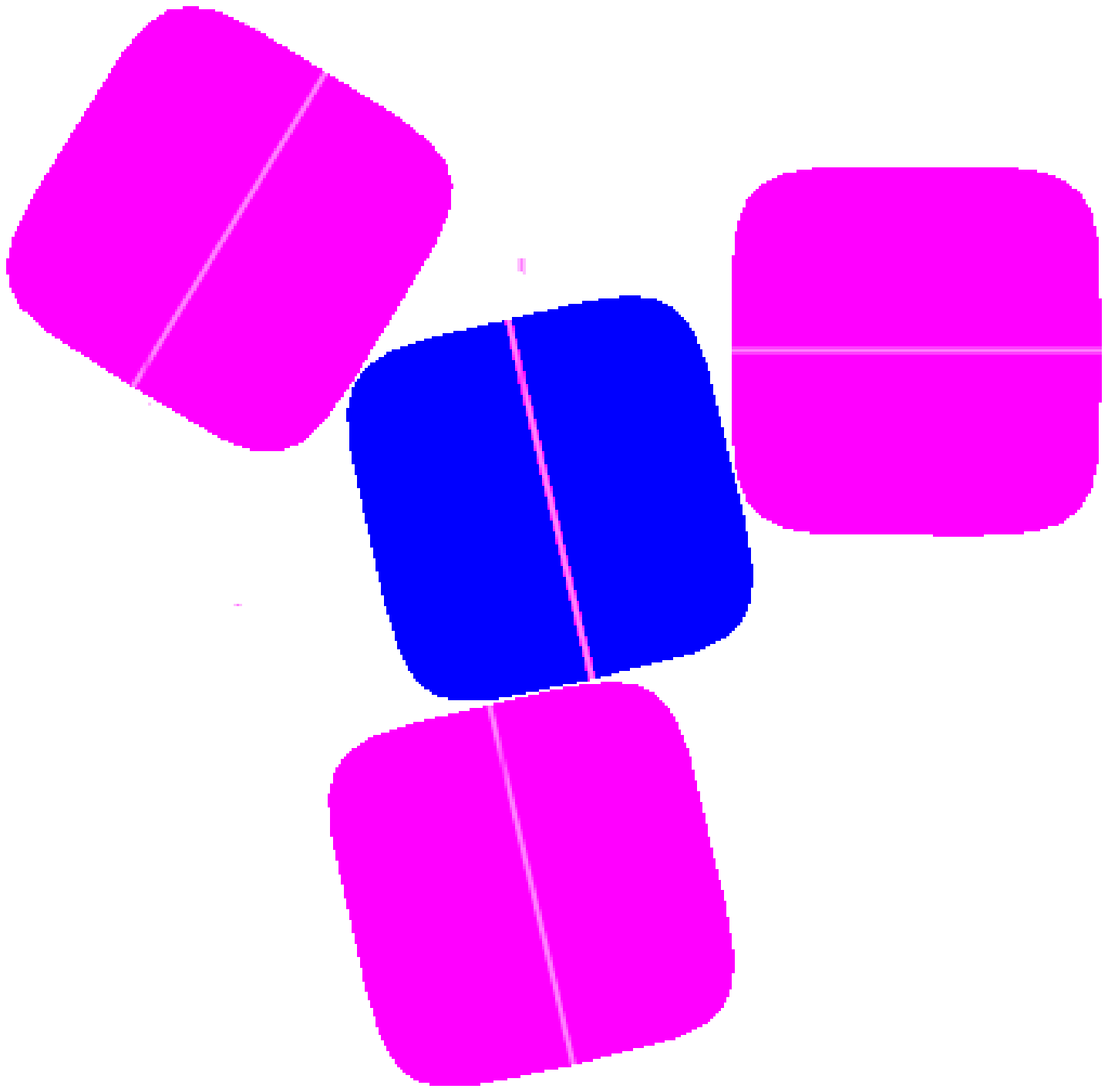} &
\includegraphics[height=4.0cm, keepaspectratio]{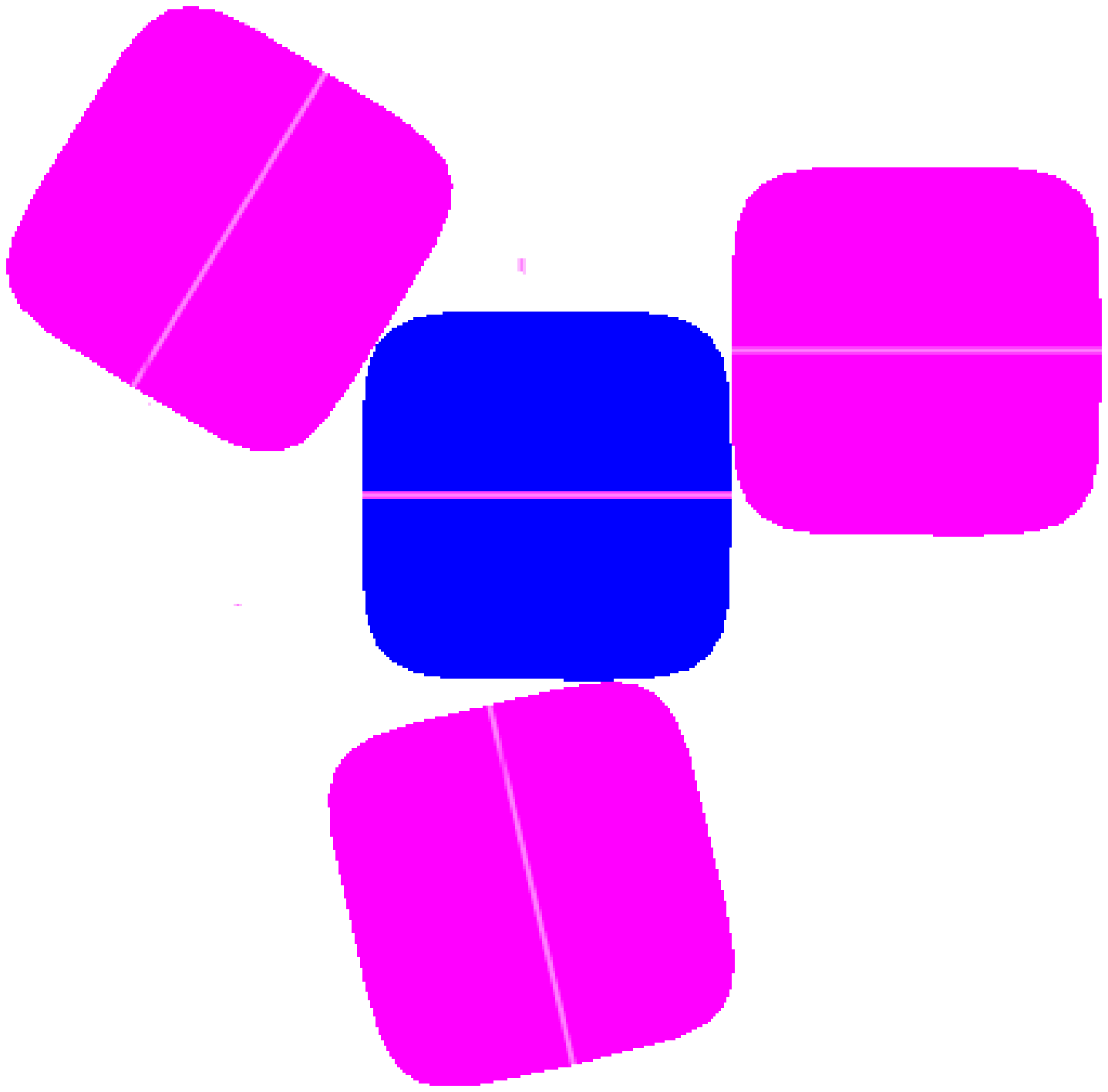} &
\includegraphics[height=4.0cm, keepaspectratio]{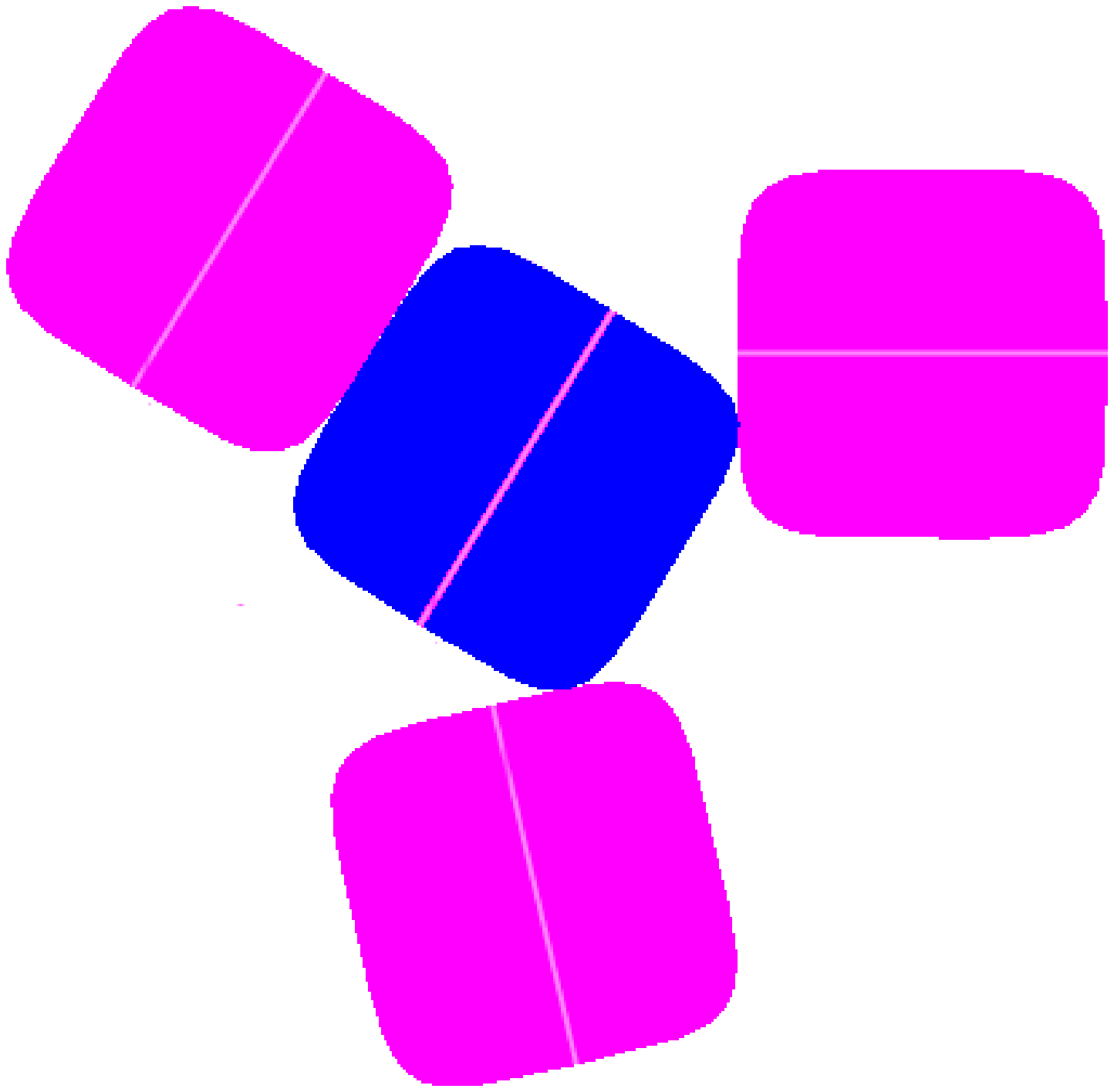} \\
\mbox{(a)} & \mbox{(b)} & \mbox{(c)}
\end{array}$
\end{center}
\caption{(color online). Nongeneric locally jammed configurations associated with three 
fixed superdisks (pink) and the trapped one (blue). In each configuration, the central superdisk 
is approximately aligned with one of its fixed neighbors to form contacts associated with 
small curvatures to block rotation.}
\label{local}
\end{figure}

We note that the aforementioned nongeneric structures (see Fig.~\ref{Generic}(b)) are not rare. 
In particular, a nonspherical particle can be rotationally jammed if it has 
neighbors that can translationally jam the particle \cite{AlexPREJam}. 
To illustrate this point, we will consider a small packing 
composed of four superdisks in two dimensions.
Now we show that one can locally jam a superdisk by three contacting neighbor superdisks. 
Translational jamming requires that the centroids of the neighbors cannot lie in the same 
semi-circle around the centroid of the central superdisk. Suppose 
a superdisk is translationally trapped (not jammed) by its three neighbors, 
whose positions and orientations are fixed. This four-particle 
configuration has four degrees of freedom: two translational and one 
rotational degrees of freedom of the trapped particle as well as 
the expansion of the particles. To obtain a jammed 
configuration, the four degrees of freedom need to be completely constrained. 
This can be achieved by the three contact 
conditions for the jammed particles and its neighbors and the 
requirement that the three inward normal vectors at the contacting points meet 
at a common point, a sufficient condition for torque balance \cite{AlexPREJam}. 
Thus, one has four independent equations for the four degrees of freedom; see the 
Appendix for details.

Figure~\ref{local} shows the nongeneric jammed configurations associated 
with three specific fixed trapping superdisks. The multiplicity of 
the configurations is due to the multiple solutions of the equations. 
The jamming configurations can be also obtained using 
the DTS algorithm, which allows the trapped 
particle to translate and rotate and allows all the particles to grow.
Although the above analysis is for local jamming (i.e., the neighbors of the central particle are fixed), 
it is reasonable to expect that collective particle rearrangements further 
facilitate the formation of nontrivial orientational correlations and thus, 
enable a larger number of nongeneric jamming configurations. Indeed, 
the numerous hypostatic jammed packings that we found from our simulations 
strengthen our argument that  nongeneric structures are not rare.


\subsection{Nonvanishing Orientational Order}

We also observe the increase of the orientational order 
(measured by $P_4$) associated with the increasing $p$ values, 
although the largest possible expansion rate $\gamma$ has been 
used to suppress the formation of orders 
(i.e., to maintain the maximal degree of randomness) \cite{exp_rate}.
As $p$ deviates from unity, the particle shape develops 
``edges'' and ``corners'' with large curvatures, which may 
not be able to block rotational unjamming motions if 
contacts occur at significantly curved regions of the particle surface. 
On the other hand, low-curvature contacts are more favorable, which is associated 
with partial alignments of the particles. The tendency of 
particle alignments to form low-curvature surface contacts 
required by jamming becomes stronger as the particle moves 
further away from the sphere point. Thus, there is also 
a competition between orientational disorder and jamming 
for packings of superdisks and superballs, resulting from 
their unique symmetry-breaking manner, which has not 
been observed in random packings of ellipsoids. 
Due to numerical difficulties, we could not use the DTS algorithm 
to study the random jammed packings of particles with extreme shapes, 
i.e., in the limit $p \rightarrow 0.5$ and $p \rightarrow \infty$. 
However, it reasonable to expect considerable 
orientational ordering in such packings.


\section{Conclusions}

In this paper, we studied the maximally random jammed packings of superdisks and superballs.
The packing densities increase dramatically and nonanalytically 
as one moves away from the circular-disk and sphere point ($p=1$) and 
the packings are hypostatic. To achieve jamming, 
the local arrangements of particles are necessarily non-trivially 
correlated and we term these structures ``nongeneric'' in light 
of the correlations. The degree of ``nongenericity'' of the packings 
is quantitatively characterized by the fraction of local structures composed of 
particles with less contacts than average. 
Moreover, we showed that such seemingly ``special'' packing configurations are not 
rare. As the anisotropy of the particles increases, 
the fraction of rattlers decreases while the minimal orientational order increases.
The novel features arising in MRJ packings of superdisks and superballs 
result from the unique manner in which rotational symmetry is broken.
This makes such packings distinctly different from other known 
MRJ packings of nonspherical particles, such ellipsoids and ellipses. 

The ability to produce dense random packings using 
superballs casts new lights on several industrial 
processes, such as sintering and ceramic formation, 
where interest exists in increasing the density of 
powder particles to be fused. If superball-like particles 
instead of spherical particles are used, the packing 
density of a randomly poured and compacted powder could 
be increased to a value surpassing that of the maximal 
sphere-packing density. We note that superdisks and superballs 
can be experimentally mass produced using current 
lithography techniques. Understanding 
the statistical thermodynamics of the jamming transition 
of superdisks and superballs, especially the role of rotational 
and translational degrees of freedom for different 
deformation parameters is a subject that merits future investigation. 
Such studies could may our understanding of the nature of glass transitions, 
since the preponderance of previous investigations have 
focused on spherical particles.


\renewcommand{\theequation}{A-\arabic{equation}} 
\setcounter{equation}{0}  
\section*{APPENDIX: EQUATIONS FOR LOCALLY JAMMED FOUR-SUPERDISK CONFIGURATIONs}

In this section, we provide the equations that determine locally jammed four-superdisk configurations 
composed of a trapped central particle and three fixed contacting neighbors (see Fig.~\ref{local}). In particular, the 
boundary of a superdisk with radius $R$ is define by 

\begin{equation}
\label{eqA01}
\left |{\frac{x_1}{R}}\right |^{2p}+\left |{\frac{x_2}{R}}\right |^{2p} = 1,
\end{equation}

\noindent which can be also expressed by the parametric equations

\begin{equation}
\label{eqA02}
\begin{array}{c}
x_1(\theta) = |\cos\theta|^{\frac{1}{p}}\cdot R\cdot sign(\cos\theta), \\
x_2(\theta) = |\sin\theta|^{\frac{1}{p}}\cdot R\cdot sign(\sin\theta),
\end{array}
\end{equation}

\noindent where $sign(x)$ gives the sign of argument $x$. Let the centroids of 
the three fixed neighbors be $(a_i, b_i)$ $(i=1, 2, 3)$, and the orientations 
be $\theta_i$. Their boundaries are then given by 

\begin{equation}
\label{eqA03}
\begin{array}{c}
x^{(i)}_1(\theta) = a_i+\cos\theta_i|\cos\theta|^{\frac{1}{p}}\cdot R\cdot sign(\cos\theta)+\sin\theta_i|\sin\theta|^{\frac{1}{p}}\cdot R\cdot sign(\sin\theta), \\
x^{(i)}_2(\theta) = b_i-\sin\theta_i|\cos\theta|^{\frac{1}{p}}\cdot R\cdot sign(\cos\theta)+\cos\theta_i|\sin\theta|^{\frac{1}{p}}\cdot R\cdot sign(\sin\theta).
\end{array}
\end{equation}

\noindent Similarly, if the centroid of the central particle is at $(a_o, b_o)$ and its 
orientation is characterized by $\theta_o$, its boundary is specified by 

\begin{equation}
\label{eqA04}
\begin{array}{c}
x^{o}_1(\theta) = a_o+\cos\theta_o|\cos\theta|^{\frac{1}{p}}\cdot R\cdot sign(\cos\theta)+\sin\theta_o|\sin\theta|^{\frac{1}{p}}\cdot R\cdot sign(\sin\theta), \\
x^{o}_2(\theta) = b_o-\sin\theta_o|\cos\theta|^{\frac{1}{p}}\cdot R\cdot sign(\cos\theta)+\cos\theta_o|\sin\theta|^{\frac{1}{p}}\cdot R\cdot sign(\sin\theta).
\end{array}
\end{equation}

\noindent Since the positions and orientations of the three neighbors are fixed, the four-particle system 
has four degrees of freedom, namely the position $(a_o, b_o)$ and orientation $\theta_o$ 
of the central particle, as well as the radius $R$ of all particles, as discussed in Sec.~III.C. 

In the jammed configuration, the central particle contacts all its three neighbors. From 
Eqs.~(\ref{eqA01}) and (\ref{eqA04}), the contact point $(x^{(i)}_{1c}, x^{(i)}_{2c})$ between 
neighbor particle $i$ and the central particle can be expressed in terms of 
$(a_o,b_o,\theta_o,R)$, which must also lie on the boundary of the neighboring particle $i$, i.e.,

\begin{equation}
\label{eqA05}
\left |{\frac{x^{(i)}_{1c}(a_o,b_o,\theta_o,R)-a_i}{R}}\right |^{2p}+\left |{\frac{x^{(i)}_{2c}(a_o,b_o,\theta_o,R)-b_i}{R}}\right |^{2p} = 1,
\end{equation}

\noindent for $i=1, 2, 3$. This leads to three equations in the variables $a_o,b_o,\theta_o$, and $R$. 
In addition, to achieve jamming the three normals at contacts must meet at 
a common point, which guarantees torque balance. The normals at contacts are along the lines

\begin{equation}
\label{eqA06}
(x_2-x^{(i)}_{2c}) = -\left .{\frac{dx^{(i)}_1/d\theta}{dx^{(i)}_2/d\theta}}\right|_{(x^{(i)}_{1c}, x^{(i)}_{2c})}(x_1-x^{(i)}_{1c}).
\end{equation}

\noindent The aforementioned torque balance condition requires that the three lines given 
by Eq.(\ref{eqA06}) must intersect at a common point. This leads to another equation in the variables $a_o,b_o,\theta_o$, and $R$.
Thus, there are four independent equations for the four degrees of freedom and $(a_o,b_o,\theta_o,R)$ 
can be completely determined for a locally jammed four-particle configuration.

\section*{ACKNOWLEDGMENTS}
The authors thank Robert  Batten and Aleksandar Donev for valuable
discussions. S. T. thanks  the Institute for Advanced Study
for its hospitality during his  stay there.
This work was supported by the Division of Mathematical Sciences
at the National Science Foundation under Award Number DMS-0804431
and by the MRSEC Program of the
National Science Foundation under Award Number DMR-0820341.


\begin{thebibliography}{31}
\bibitem{supra}
Supramolecular chemistry deals with the chemistry and collective 
behavior of molecular building blocks that are organized on
large length scales (relative to molecular sizes)
with long-range order.
\bibitem{aspect}
The aspect ratio of a ellipsoid $(x/a)^2+(y/b)^2+(z/c)^2=1$ is defined as the ratio $c/a$.
\bibitem{ftn_degenerate}
In a previous study of jammed random ellipsoid packings reported in Ref. \cite{AlexPREJam}, 
nontrivially correlated local structures were called ``degenerate.'' In light 
of the fact that ``degenerate'' has a variety of different meanings,  
we use instead the term ``nongeneric'' to avoid any possible confusion.
\bibitem{shrinkage}
Theoretically, the particle should be shrunk by an infinitesimal amount and one 
should check whether there 
exist nontrivial unjamming particle motions that preserve the nonoverlapping conditions up to 
the same order of magnitude as the relative amount of shrinkage. 
To implement this method numerically for the system sizes
that have been examined, we shrink the particles by a finite 
small amount such that the sizes of inter-particle gaps increase from $10^{-6}$ to $10^{-4}$ 
of the particle diameter.
\bibitem{slope2d}
The numerically computed right and left slope of $\phi(p)$ for binary 
superdisks at $p=1$ are $a_{+} \approx 0.135$ and $a_{-} \approx -0.083$, respectively.
\bibitem{slope3d}
The numerically computed right and left slope of $\phi(p)$ for monodispersed 
superballs at $p=1$ are $a_{+} \approx 0.229$ and $a_{-} \approx -0.113$, respectively.
\bibitem{rotation}
The fact that the superdisks and superballs 
can significantly explore their configurational 
space associated with the rotational degrees of freedom, i.e., they can 
rotate significantly, near the jamming point is manifested as long-period 
oscillations of the pressure of the system, which is 
also observed for ellipsoid systems.
\bibitem{ellipse}
In this sense, it is similar to the ellipse packings in two dimensions, 
i.e., the isostatic packing requires six contacts per particles, 
which can only be realized with translational crystallization.
\bibitem{exp_rate}
Expansion rates $\gamma$ within the range $0.1-0.5$ were employed 
to suppress the formation of order in the superdisk and 
superball packings; larger $\gamma$ would cause numerical 
instability of the DTS algorithm. Note that for ellipsoids, 
$\gamma \sim 0.05$ is sufficient to 
produce random packings with vanishing orientational order \cite{AlexSci,AlexPREJam}.
\end{thebibliography}
\end{document}